\documentclass[conference]{IEEEtran}
\usepackage[table,xcdraw]{xcolor}
\usepackage{hyperref}
\usepackage{makecell} 
\IEEEoverridecommandlockouts
\usepackage{cite}
\usepackage{amsmath,amssymb,amsfonts}
\usepackage{algorithmic}
\usepackage{graphicx}
\usepackage{textcomp}
\usepackage{xcolor}
\def\BibTeX{{\rm B\kern-.05em{\sc i\kern-.025em b}\kern-.08em
    T\kern-.1667em\lower.7ex\hbox{E}\kern-.125emX}}
\begin{document}

\title{Toward Effective AI Governance: A Review of Principles}
\author{
\IEEEauthorblockN{
Danilo Ribeiro,
Thayssa Rocha,
Gustavo Pinto,
Bruno Cartaxo,
Marcelo Amaral,
Nicole Davila,
Ana Camargo
}
\IEEEauthorblockA{
Zup Innovation\\
Brazil \\
\{danilo.ribeiro, gustavo.pinto, bruno.cartaxo, marcelo.amaral, nicole.davila,\\ thayssa.rocha, ana.camargo\}@zup.com.br
}
}


\maketitle

\begin{abstract}
\textbf{Background:} Artificial Intelligence (AI) governance is the practice of establishing frameworks, policies, and procedures to ensure the responsible, ethical, and safe development and deployment of AI systems. Although AI governance is a core pillar of Responsible AI, current literature still lacks synthesis across such governance frameworks and practices. \textbf{Objective:} To identify which frameworks, principles, mechanisms, and stakeholder roles are emphasized in secondary literature on AI governance.
\textbf{Method:} We conducted a rapid tertiary review of nine peer-reviewed secondary studies from IEEE and ACM (2020–2024), using structured inclusion criteria and thematic-semantic synthesis. \textbf{Results:} The most cited frameworks include the EU AI Act and NIST RMF; transparency and accountability are the most common principles. Few reviews detail actionable governance mechanisms or stakeholder strategies. \textbf{Conclusion:} The review consolidates key directions in AI governance and highlights gaps in empirical validation and inclusivity. Findings inform both academic inquiry and practical adoption in organizations.
\end{abstract}

\begin{IEEEkeywords}
Responsible Artificial Intelligence, Tertiary Review, AI Ethics
\end{IEEEkeywords}

\section{Introduction}

The increasing deployment of Artificial Intelligence (AI) systems across critical domains, such as healthcare, finance, public administration, has raised urgent concerns about their ethical, legal, and social implications~\cite{floridi2018ai4people}. In response, the notion of \textit{Responsible AI} (RAI) has emerged as a multidimensional paradigm that seeks to ensure fairness, transparency, accountability, privacy, safety, and human oversight in the development and use of AI technologies~\cite{floridi2018ai4people}.

A key component of RAI is the establishment of effective \textbf{governance mechanisms}, which encompass regulatory frameworks, organizational structures, internal processes, and stakeholder engagement strategies that ensure AI systems are trustworthy and aligned with societal values~\cite{raji2020closing}. In this context, Governance is not limited to compliance with legal norms. However, it includes the design of institutional arrangements that operationalize ethical principles in real-world settings—particularly in technology companies, in which AI systems are developed and deployed at scale~\cite{raji2020closing}.

Over the past years, academic interest in AI governance has intensified~\cite{aiindex2023, feldman2024diffusion}. Numerous systematic reviews and scoping studies have been published, covering ethical principles, explainability techniques, regulatory developments, and organizational practices. However, the growing number of such reviews, we believe it is time to better understand important key points regarding, for instance, how governance has been conceptualized, which frameworks and practices are emerging, and how stakeholder roles are being addressed in the literature.

This paper presents a \textbf{rapid tertiary review} of the secondary literature on Responsible AI governance to address this gap. We call it \emph{rapid} because it uses elements of a Rapid Review~\cite{cartaxo2020,Cartaxo2018} (more in Section~\ref{sec:methodology}. By synthesizing findings from systematic reviews, scoping studies, mapping studies, and multivocal literature reviews published between 2020 and 2024, we aim to provide an integrative perspective on RAI's state of the art. Our review focuses not only on theoretical framings but also on practical mechanisms and relevant recommendations to organizations, particularly those involved in the development and application of AI technologies.

We seek to answer four research questions: (1) What frameworks of AI governance are most frequently cited in the literature? (2) Which principles are emphasized across secondary studies? (3) What organizational structures and governance mechanisms are recommended? and (4) How are stakeholders involved and represented in governance discussions? By addressing these questions, we aim to support both researchers and practitioners in designing governance strategies that are theoretically grounded and practically actionable.

The rapid tertiary review of AI governance literature yielded three primary insights:

\begin{itemize}
    \item \textbf{Dominance of High-Level Regulatory Frameworks with Operational Gaps:} The studies frequently reference established frameworks such as the EU AI Act and NIST RMF. However, there is a notable dearth of actionable governance mechanisms and stakeholder engagement strategies detailed within these secondary reviews. This indicates a significant gap between prescriptive regulatory guidance and concrete, empirically validated implementation practices.

    \item \textbf{Prevalence of Transparency and Accountability Principles:} Transparency and accountability consistently emerge as the most emphasized governance principles across the analyzed literature. These are often discussed alongside other core tenets like fairness, explainability, and privacy. This highlights a shared conceptual foundation for responsible AI.

    \item \textbf{Call for Empirical Validation and Enhanced Inclusivity:} The review underscores a critical need for empirical validation of proposed AI governance practices. Furthermore, it identifies a deficiency in the literature regarding the detailed exploration and effective integration of diverse stakeholder perspectives, particularly those of underrepresented groups. This suggests a requirement for future research to move beyond conceptual discussions to real-world impact assessments and more inclusive governance models.
\end{itemize}

The remainder of this paper is organized as follows. Section~\ref{sec:methodology} describes the methodology adopted for this rapid tertiary review, including the search strategy, inclusion criteria, and synthesis approach. Section~\ref{sec:results} presents the main findings, structured around the four research questions. Section~\ref{sec:discussion} discusses the implications of our results for industry, society, and future research. Section~\ref{sec:validity} addresses threats to validity. Section~\ref{sec:conclusion} concludes the paper with final remarks and directions for future work.

\section{Methodology}
\label{sec:methodology}

This study adopts a rapid tertiary review methodology, which merges the scope of tertiary evidence synthesis with the time-efficiency of a rapid review~\cite{cartaxo2020rapid}. This approach is particularly suitable for emerging research topics, like AI Governance, where timely insights are prioritized over exhaustive coverage. To ensure efficiency, the review was conducted under streamlined conditions: only the first author performed the literature screening and data extraction and the search was limited to two well-established digital libraries—IEEE Xplore and the ACM Digital Library. The review was performed in May 2025.

Then, we belive that tertiary reviews aggregate and analyze existing secondary studies, such as systematic reviews and scoping reviews, while rapid reviews adopt streamlined procedures to deliver timely insights under practical constraints. This hybrid approach is especially useful in dynamic and multidisciplinary domains such as Responsible AI governance.

\subsection{Research Questions}

The aim of this study is to investigate how governance in RAI has been addressed in the secondary literature. Our focus is particularly on frameworks, governance principles, stakeholder engagement, and organizational mechanisms relevant to technology companies. The following research questions were defined to guide the analysis:

\begin{itemize}
    \item \textbf{RQ1: What are the main AI governance frameworks discussed in secondary reviews?} \\
    \textit{Justification:} This question seeks to identify which normative and regulatory frameworks (e.g., EU AI Act, NIST RMF) are most cited in review literature, revealing their influence on research, policy, and organizational adoption of Responsible AI.

    \item \textbf{RQ2: Which governance principles (e.g., transparency, accountability, auditability) are most frequently addressed?} \\
    \textit{Justification:} By mapping which principles are emphasized—such as fairness, privacy, or explainability—this question reveals ethical priorities and possible gaps or tensions in the governance discourse across different contexts.

    \item \textbf{RQ3: What organizational structures and internal governance mechanisms are identified as good practices?} \\
    \textit{Justification:} Operationalizing AI governance requires institutional arrangements. This question examines the organizational practices (e.g., AI ethics committees, algorithmic audits, documentation protocols) highlighted as effective for implementing Responsible AI.

    \item \textbf{RQ4: What is the role of stakeholders (e.g., regulators, developers, users, citizens) in AI governance?} \\
    \textit{Justification:} AI governance is a sociotechnical challenge. This question investigates how reviews classify and engage with different stakeholders, including their influence, responsibilities, and representation within governance frameworks.
\end{itemize}

\subsection{Search Strategy}

The search was conducted in April and May 2025 using two major digital libraries: IEEE Xplore and ACM Digital Library. The following query string was applied to search titles and abstracts:

\begin{quote}
(``AI governance'' OR ``AI compliance'') AND \\
(``systematic review'' OR ``scoping review'' OR ``literature review'' OR ``mapping study'' OR ``meta-analysis'' OR ``research synthesis'')
\end{quote}

Filters were applied to restrict results to English-language, peer-reviewed articles published between 2020 and 2024.

\subsection{Inclusion and Exclusion Criteria}

\textbf{Inclusion criteria:}
\begin{itemize}
    \item The article is a secondary study (e.g., systematic review, scoping review, mapping study, multivocal review, or meta-analysis).
    \item It explicitly addresses Responsible AI and governance-related topics.
    \item It includes practical governance recommendations, mechanisms, or frameworks relevant to technology companies or organizations implementing AI.
    \item It is peer-reviewed, written in English, and published between 2020 and 2024.
    \item The full text is accessible for analysis.
\end{itemize}

\textbf{Exclusion criteria:}
\begin{itemize}
    \item The article is a primary study, opinion piece, or theoretical essay without review methodology.
    \item It does not include governance-related content (e.g., discusses only ethical theory without implementation).
    \item It lacks practical relevance to organizations or industries using AI.
    \item It was published before 2020, is not peer-reviewed, not written in English, or is not available in full text.
\end{itemize}

\subsection{Study Selection}

A total of 55 articles were retrieved across IEEE Xplore and ACM Digital Library. After applying the inclusion and exclusion criteria through manual screening of titles, abstracts, and full texts, 9 articles were selected for detailed analysis. These studies met all methodological and thematic requirements of our study, and presented governance solutions, classifications, or organizational mechanisms applicable to real-world Responsible AI adoption.

\subsection{Data Extraction and Analysis}

Data were extracted into a structured spreadsheet that captured bibliographic information, type of review, year of publication, governance themes discussed, and relevance to each research question. A classification was also applied based on six key RAI governance pillars, inspired in recent literature on the topic~\cite{pilares,floridi2018ai4people}: fairness, transparency, privacy and security, sustainability, accountability, and explainability.

The complete pillars description, and data (selection and extraction) sheets are available for open science~\footnote{https://anonymous.4open.science/r/tertiaryreview/}.

\subsection{Synthesis Approach}

We used a thematic synthesis approach, following the recommended steps: extract data, code data, translate code into themes, and create a model of higher-order themes~\cite{CruzesDyba2011}. Our analysis combined descriptive mapping with interpretative synthesis. Whenever possible, original excerpts from the studies were preserved and integrated into the discussion to ensure analytical transparency and maintain fidelity to the reviewed literature. Special attention was given to identifying actionable governance mechanisms and practices recommended for implementation in technology companies.

\subsection{Limitations}

This rapid tertiary review prioritizes coverage and insight over exhaustiveness. The scope was limited to two digital libraries and a five-year publication window. The quality and depth of individual secondary studies also varied. However, we mitigated these risks through strict inclusion criteria, manual validation, and methodological triangulation (e.g., thematic and semantic classification).

\section{Results}
\label{sec:results}
This section presents the synthesis of findings from the selected secondary reviews, structured around four research questions defined for this tertiary study.
\subsection{General Results}
Table~\ref{tab:included_studies} presents the nine  secondary studies included in this rapid tertiary review. All selected articles were published between 2020 and 2024 in peer-reviewed venues, including conferences (e.g., FAccT, ICSR) and indexed journals (e.g., IEEE Access, ACM Computing Surveys).

The studies cover diverse governance-related themes, such as explainability, stakeholder engagement, internal accountability, and metrics for trustworthy AI. The combination of conceptual and practical contributions across the selected articles provides a robust foundation for answering the research questions proposed in this review.
\begin{table}[ht]
\centering
\caption{Included studies in the review.}
\label{tab:included_studies}
\resizebox{\columnwidth}{!}{%
\begin{tabular}{|l|c|}
\hline
\textbf{Title} & \textbf{Year} \\
\hline
\makecell[l]{Towards a Privacy and Security-Aware Framework for Ethical AI:\\ Guiding the Development and Assessment of AI Systems~\cite{abelnica2024governance}} & 2024 \\
\hline
\makecell[l]{Typology of Risks of Generative Text-to-Image Models~\cite{bird2023typology}} & 2023 \\
\hline
\makecell[l]{Towards a Responsible AI Metrics Catalogue:cA Collection of \\ Metrics for AI Accountability~\cite{korobenko2024privacy}} & 2024 \\
\hline
\makecell[l]{Developing an Ethical Regulatory Framework  for Artificial\\ Intelligence: Integrating Systematic Review, Thematic Analysis,\\ and Multidisciplinary Theories~\cite{deshpande2022stakeholders}} & 2024 \\
\hline
\makecell[l]{IT Governance in the Artificial Intelligence Age: Trends\\ and Practices~\cite{abelnica2024governance}} & 2024 \\
\hline
\makecell[l]{A Roadmap of Explainable Artificial Intelligence: Explain to Whom, \\When, What and How?~\cite{wang2024roadmap}} & 2024 \\
\hline
\makecell[l]{A Systematic Literature Review on AI-Based Recommendation \\ Systems and Their Ethical Considerations~\cite{masciari2024recommendation}} & 2024 \\
\hline
\makecell[l]{Responsible AI Pattern Catalogue: A Collection of Best Practices \\ for AI Governance and Engineering~\cite{lu2023responsible}} & 2023 \\
\hline
\makecell[l]{Responsible AI Systems: Who are the Stakeholders?~\cite{raji2020closing}} & 2024 \\
\hline

\end{tabular}
}
\end{table}

\subsection*{RQ1: What are the main AI governance frameworks discussed in secondary reviews?}

The reviewed studies present a variety of frameworks aimed at operationalizing Responsible AI. One prominent example is the Responsible AI Pattern Catalogue, which organizes practices into three categories: multi-level governance patterns, trustworthy process patterns, and responsible-by-design product patterns. These patterns were identified through a multivocal literature review and are intended to support the systemic implementation of Responsible AI throughout the AI lifecycle~\cite{lu2023responsible}.

Another contribution is the Responsible AI Metrics Catalogue, which focuses specifically on accountability. It proposes a structured set of metrics organized into process, resource, and product categories. These metrics are designed to fill existing gaps in practical guidance for operationalizing AI accountability, particularly in the context of generative AI~\cite{xia2024metrics}.

A systematic review presents a conceptual Privacy and Security-Aware Framework for Ethical AI, structured around four dimensions: data, technology, people, and process. This framework provides a foundation for the development and evaluation of AI systems with integrated privacy and security concerns~\cite{korobenko2024privacy}.

Other studies discuss initiatives and frameworks such as AI Verify (Singapore), the EU's capAI project, the NIST AI Risk Management Framework, the EU Trustworthy AI Assessment List, the NSW AI Assurance Framework (Australia), and Microsoft’s Responsible AI Impact Assessment Template~\cite{xia2024metrics,lu2023responsible}.

International guidelines developed by governmental and intergovernmental bodies, such as the OECD, G7, G20, and national governments (EU, US, UK, Canada, Australia), are also widely referenced. Examples include the EC-HLEG AI guidelines, the Montreal Declaration, and the Beijing AI Principles~\cite{abelnica2024governance,lu2023responsible,xia2024metrics}.

Contributions from technology companies such as Microsoft, Google, and IBM have also been cited in the secondary reviews, particularly in relation to their frameworks and assessment templates for Responsible AI~\cite{lu2023responsible,xia2024metrics,korobenko2024privacy}.

Professional organizations, including the ACM and IEEE, are frequently referenced, especially for their ethical guidelines and design standards for intelligent systems~\cite{abelnica2024governance,lu2023responsible}.

Additionally, research institutes such as the Alan Turing Institute are recognized for their significant contributions to AI ethics and governance research~\cite{abelnica2024governance}.

\subsection*{RQ2: Which governance principles (e.g., transparency, accountability, auditability) are most frequently addressed?}

The reviews consistently emphasize a core set of ethical principles, often reflecting national and international guidelines. These include human-centered values, social and environmental well-being, fairness, privacy and security, reliability and safety, transparency, explainability, contestability, and accountability~\cite{lu2023responsible}.

Accountability is a central theme in the Responsible AI Metrics Catalogue. It is defined through three complementary elements: responsibility, auditability, and redressability. These components are essential for transparent and auditable decision-making, building public trust, and complying with emerging regulations~\cite{xia2024metrics}.

The emphasis on privacy and security is especially notable in the Privacy and Security-Aware Framework, which highlights the need for an integrated approach. The study finds that privacy is widely addressed in the literature, whereas security is less frequently discussed~\cite{korobenko2024privacy}.

Explainability and transparency are key principles in reviews dedicated to explainable AI. One such study outlines their importance in addressing the lack of interpretability in AI systems and in meeting regulatory requirements~\cite{wang2024roadmap}.

Across all studies, transparency and privacy are the most frequently cited principles, followed by fairness, accountability, explainability, autonomy, responsibility, and safety~\cite{lu2023responsible,xia2024metrics}.

\subsection*{RQ3: What organizational structures and internal governance mechanisms are identified as good practices?}

The reviews identify a variety of internal governance mechanisms considered effective for supporting Responsible AI. The Responsible AI Pattern Catalogue emphasizes the importance of multi-level governance structures involving the industry, the organization, and technical teams~\cite{lu2023responsible}.

The establishment of AI governance committees is described as an effective practice. These committees should include professionals from diverse areas and strategic leadership to ensure ethical oversight across the entire lifecycle of AI systems~\cite{lu2023responsible,xia2024metrics}.

The literature also suggests establishing formal processes for ethical oversight, compliance verification, and incident response. It recommends competence assessments tailored to different roles in the organization, aligned with industry standards~\cite{xia2024metrics}.

Responsible AI maturity models are presented as useful tools to evaluate and improve organizational capabilities in AI governance. Certification mechanisms are also proposed to demonstrate compliance with ethical standards~\cite{lu2023responsible}.

Standardized reporting is highlighted as a necessary practice for transparency in communication with stakeholders. This includes disclosing when AI is being used and explaining its purpose and design~\cite{lu2023responsible,xia2024metrics}.

Some studies emphasize internal changes in organizations, such as the creation of AI ethics teams, adoption of internal governance guidelines, and the training of developers and engineers in ethics and human rights~\cite{lu2023responsible,korobenko2024privacy}.

\subsection*{RQ4: How do secondary reviews discuss the role of stakeholders (e.g., regulators, developers, users, citizens) in AI governance?}

The stakeholders role in Responsible AI governance is discussed extensively across the reviews. The Responsible AI Pattern Catalogue categorizes stakeholders into three levels: industry, organizational, and team. At the industry level, policymakers and regulators act as enablers, while technology producers and procurers are key affected parties. At the organizational level, managers are responsible for governance structures, influencing employees, users, and individuals impacted by AI. Development teams are directly involved in implementing Responsible AI in practice and product design~\cite{lu2023responsible}.

Several frameworks highlight the collaborative nature of Responsible AI, requiring engagement from policymakers, developers, end-users, and civil society. The Privacy and Security-Aware Framework, for example, is designed to support public institutions, private companies, and academia in addressing shared concerns~\cite{korobenko2024privacy}.

In the context of explainable AI, different stakeholder groups, such as AI experts, decision-makers, regulators, and users, are considered. One study identifies nine distinct stakeholder types with varying needs for explanation~\cite{wang2024roadmap}.

Other reviews aim to support AI system builders by helping them select appropriate governance guidelines. These builders include technical professionals but also interact with business executives, legal advisors, and policymakers~\cite{xia2024metrics}.

The literature also acknowledges that AI governance involves addressing the needs of underrepresented and vulnerable populations, and that trade-offs must be managed across individual, organizational, and systemic levels~\cite{lu2023responsible,xia2024metrics}.

\section{Discussion}
\label{sec:discussion}

The results of this tertiary review indicate a rapidly evolving landscape in the field of AI governance, shaped by both international regulatory initiatives and industry-driven practices. While several frameworks—such as the EU AI Act, the NIST AI RMF, and various pattern catalogs—are widely referenced, the literature shows fragmentation in how governance is conceptualized and applied. The distinction between abstract principles and their practical operationalization remains a recurring challenge.

Notably, although many reviews emphasize the importance of principles like transparency, accountability, and explainability, few go beyond descriptive mappings to provide critical analysis of how these principles interact or conflict in organizational settings. Furthermore, while stakeholder involvement is frequently acknowledged, the literature lacks depth in evaluating the effectiveness of participatory governance approaches, especially in contexts involving marginalized or underrepresented groups.

Importantly, only a subset of the reviewed studies provided concrete, organization-level governance mechanisms such as audit procedures, ethics committees, or risk management practices tailored to technology companies. This indicates a persistent gap between ethical aspirations and the availability of actionable strategies to guide the real-world implementation of Responsible AI.

We contextualized our findings on three different perspectives that are detailed below: implications for industry, society and research.
\subsection{Implications for Industry}

This review provides technology companies and AI product teams with a consolidated overview of governance mechanisms, frameworks, and practices that have been discussed and recommended in the literature. By identifying the most prominent principles (such as transparency, accountability, and fairness) and their associated implementation strategies (such as algorithmic audits and ethics committees), the review offers practical insights that can inform the development of internal Responsible AI policies. Furthermore, the categorization of stakeholder roles and organizational practices may assist in aligning cross-functional responsibilities (e.g., legal, engineering, ethics) within AI initiatives.

\subsection{Implications for Society}

The findings underscore the importance of governance structures that not only ensure legal compliance but also proactively mitigate societal risks, such as bias, discrimination, opacity, and exclusion in AI systems. The emphasis on stakeholder involvement highlights the need for more inclusive governance strategies that recognize and respond to the concerns of historically underrepresented or vulnerable populations. For policy-makers and civil society organizations, this review serves as a resource to understand how institutional and technical dimensions of AI governance are being addressed in academic and industry debates.

\subsection{Implications for Research}

This tertiary review synthesizes fragmented knowledge across multiple reviews and discloses emerging patterns, redundancies, and gaps. It contributes to Responsible AI research by mapping the literature not only by principle or framework, but also by applicability to real-world governance contexts. The study highlights the need for more empirical validation of governance practices, greater attention to organizational dynamics in AI ethics, and methodological consistency across review studies. Future research should explore interdisciplinary approaches and develop metrics to evaluate the effectiveness and fairness of governance structures in operational environments.
\section{Threats to Validity}
\label{sec:validity}

Several limitations may affect the validity of this rapid tertiary review. Following established guidelines for systematic and rapid reviews, we outline potential threats according to four dimensions: construct, internal, external, and descriptive validity.

\textbf{Construct validity} refers to the adequacy of the concepts captured by our research questions and selection criteria. Although we focused on governance in Responsible AI, the term “governance” is used inconsistently across the literature.

\textbf{Internal validity} concerns the process of study selection and data extraction. Despite applying clear inclusion and exclusion criteria, the classification of studies into governance principles and practices involved human judgment. To reduce bias, we performed manual validation. Still, some interpretations may reflect subjective alignment.

\textbf{External validity} relates to the generalizability of the findings. Our review was limited to publications from IEEE and ACM between 2020 and 2024. Although these databases are highly reputable in computing and AI, they may not fully represent the legal, sociopolitical, or multidisciplinary dimensions of AI governance found in other domains (e.g., public policy, law, or HCI). Therefore, the findings should be interpreted as reflective of the technical computing community’s perspective.

\textbf{Descriptive validity} pertains to the accuracy and completeness of reporting. We preserved excerpts from the original reviews to support transparency and traceability. However, secondary reviews sometimes lack clarity or depth in their own reporting, which may have affected our ability to fully extract or categorize content.

\section{Conclusion}
\label{sec:conclusion}

This rapid tertiary review synthesized evidence from nine secondary studies published between 2020 and 2024 on Responsible AI governance. We mapped the most referenced frameworks, identified core principles prioritized in governance discourse, highlighted recurring organizational practices, and analyzed how stakeholders are framed in governance strategies.

Our findings reinforce the importance of integrating ethical principles into formal governance structures that extend beyond compliance. The literature suggests growing convergence around certain best practices, such as ethics review boards, algorithmic auditing, and documentation practices. However, further research is needed to evaluate these practices empirically and to address stakeholder asymmetries, especially in global and industrially diverse contexts.

This study contributes by offering an integrative overview for researchers and practitioners seeking to understand how AI governance is being conceptualized, recommended, and implemented in organizational environments. It also reinforces the value of tertiary reviews in consolidating fragmented insights across a rapidly expanding field.

\section*{Acknowledgment}

The authors would like to thank the contributors of the secondary studies analyzed in this review. We also acknowledge the use of OpenAI’s ChatGPT model in supporting the academic writing process. ChatGPT was used specifically for English translation and the stylistic refinement of sections, under the authors’ full supervision. The responsibility for all content, structure, and interpretation remains solely with the authors.

\bibliography{referencia} 
\end{document}